\documentclass[sigconf,10pt]{acmart}
\AtBeginDocument{%
  }

\acmISBN{978-1-4503-XXXX-X/2018/06}

\usepackage{multicol}
\usepackage{multirow}
\usepackage{appendix}
\usepackage{caption}
\usepackage{subcaption}
\usepackage[graphicx]{realboxes}
\usepackage{booktabs}
\usepackage{url, pifont}
\usepackage{makecell}
\usepackage{amsmath,bbm}
\usepackage{appendix}

\newcommand{\eg}{\textit{e}.\textit{g}.}
\def\etc{\emph{etc}}

\copyrightyear{2025}
\acmYear{2025}
\setcopyright{rightsretained}
\acmConference[ACM MOBICOM '25]{The 31st Annual International Conference on Mobile Computing and Networking}{November 4--8, 2025}{Hong Kong, China}
\acmBooktitle{The 31st Annual International Conference on Mobile Computing and Networking (ACM MOBICOM '25), November 4--8, 2025, Hong Kong, China}\acmDOI{10.1145/3680207.3765656}
\acmISBN{979-8-4007-1129-9/2025/11}

\begin{document}

\title{Automatically Generating High-Precision Simulated Road Networking in Traffic Scenario}

\author{Liang Xie}
\email{{LXie5201@outlook.com}}
\affiliation{%
\institution{School of Computer Science and Technology, Guangdong University of
Technology}
\city{Guangzhou}
\country{China}
}
\affiliation{%
\institution{Shenzhen Graduate School, Peking University}
\city{Shenzhen}
\country{China}
}

\author{Wenke Huang}
\email{hwk727@163.com}
\affiliation{
  \institution{Peng Cheng Laboratory}
 \city{Shenzhen}
\country{China}
}



\renewcommand{\shortauthors}{Liang Xie, Wenke Huang}

\begin{abstract}
      Existing lane-level simulation road network generation is labor-intensive, resource-demanding, and costly due to the need for large-scale data collection and manual post-editing. To overcome these limitations, we propose automatically generating high-precision simulated road networks in traffic scenario, an efficient and fully automated solution. 
    Initially, real-world road street view data is collected through open-source street view map platforms, and a large-scale street view lane line dataset is constructed to provide a robust foundation for subsequent analysis. Next, an end-to-end lane line detection approach based on deep learning is designed, where a neural network model is trained to accurately detect the number and spatial distribution of lane lines in street view images, enabling automated extraction of lane information. Subsequently, by integrating coordinate transformation and map matching algorithms, the extracted lane information from street views is fused with the foundational road topology obtained from open-source map service platforms, resulting in the generation of a high-precision lane-level simulation road network. This method significantly reduces the costs associated with data collection and manual editing while enhancing the efficiency and accuracy of simulation road network generation. It provides reliable data support for urban traffic simulation, autonomous driving navigation, and the development of intelligent transportation systems, offering a novel technical pathway for the automated modeling of large-scale urban road networks.
\end{abstract}




\maketitle

\section{Introduction}



\begin{figure}[!t]
    \center
    \includegraphics[width=3.4in]{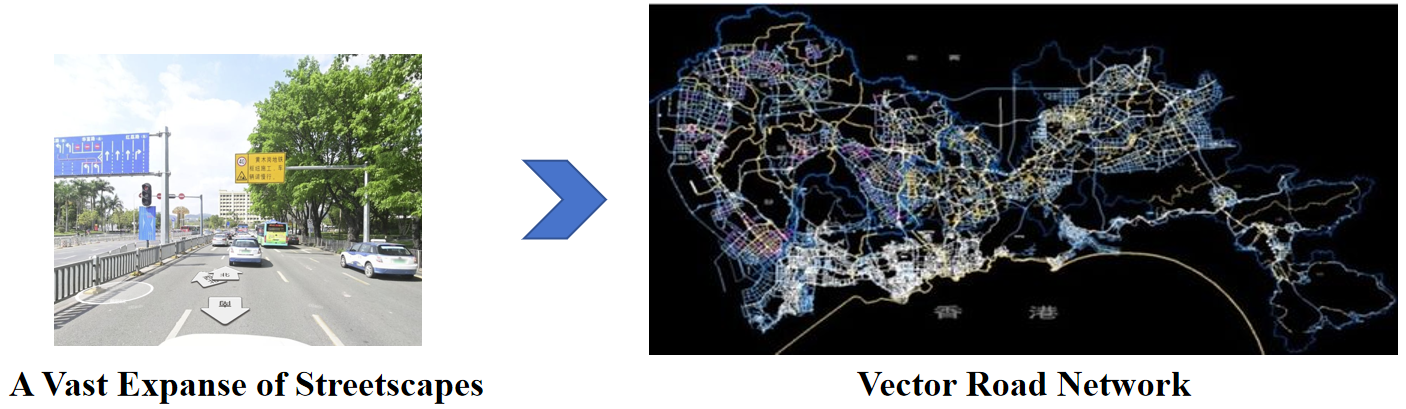}
    \caption{Can a road network be reconstructed from vast amounts of street view data?} 
    \label{fig:road_network}
\end{figure}

\begin{figure*}[!t]
    \centering
    \includegraphics[scale=0.55]{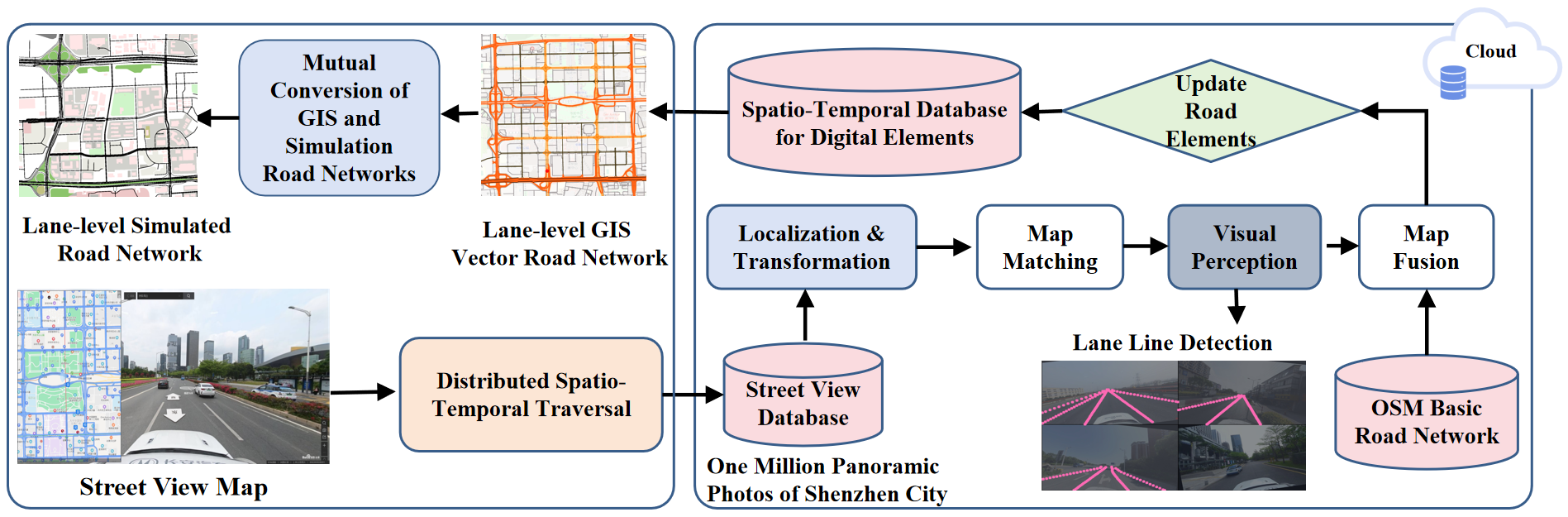}
    \caption{Automatic generation of lane-level urban simulation road networks based on street view.} 
    \label{fig: framework}
\end{figure*} 
With the rapid pace of urbanization, the expansion of urban road networks and the dramatic surge in traffic volume have led to an exponential increase in the computational demands of city-scale traffic simulation. Accurately simulating and evaluating the operational state of road networks, while conducting quantitative analysis and feedback adjustments, has become a critical yet challenging aspect of advancing technologies such as network signal control evolution and dynamic traffic assignment. Traffic simulation~\cite{nguyen2021overview,zhong2023guided}, as an effective approach to modeling the interaction between vehicles and road networks, serves as an ideal tool for fine-grained design and quantitative evaluation. At the core of traffic simulation models lies the traffic simulation road network, where the acquisition and modeling of the underlying road network represent a pivotal but challenging task. 	
Although several mainstream traffic simulation systems are available in domestic and international markets, and many large and medium-sized cities have established simulation-based decision-making platforms, these systems and platforms are often constrained by the limitations of existing simulation software, requiring manual construction of simulation road networks, which is both time-consuming and labor-intensive~\cite{li2024choose}.

\begin{figure}[!t]
    \center
    \includegraphics[width=3.4in]{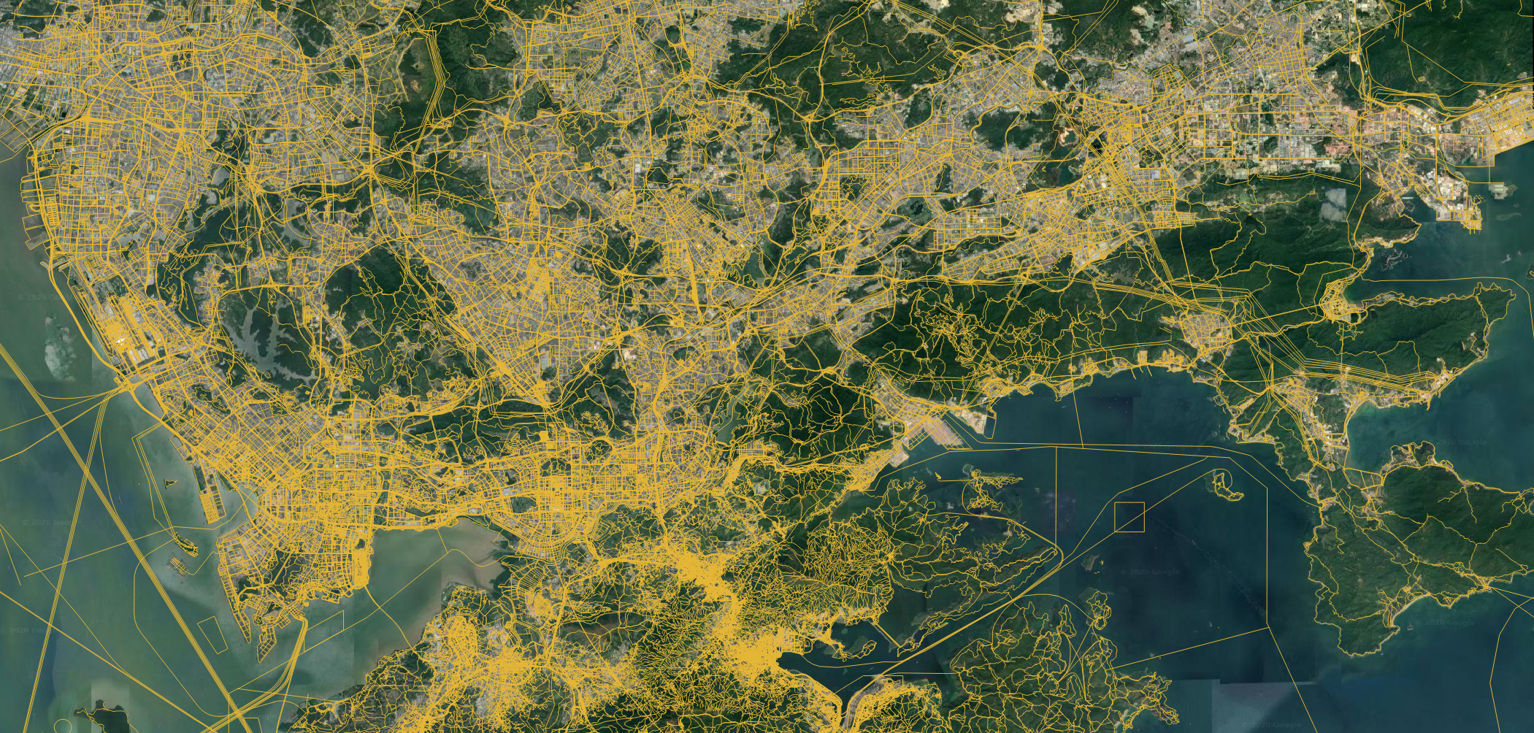}
    \caption{The base vector map data is sourced from OpenStreetMap (OSM). It is crowdsourced, open-source, and provides nationwide coverage.} 
    \label{fig:road_network1}
\end{figure}
	
Meanwhile, automated map generation~\cite{tang2022automatic,quan2025large,teramoto2025automated,fang2025scan} plays a crucial role in urban services and location-based applications (LBS), offering an efficient solution to the labor-intensive and time-consuming process of manual map creation while ensuring exceptional accuracy. 
Current research primarily relies on remote sensing image or vehicle trajectory~\cite{zha2023survey} data that can adequately reflect road network structures to generate maps. However, the reliance on single data sources limits the accuracy and completeness of the resulting maps. By effectively integrating remote sensing image with street view data and leveraging their complementary strengths, the quality of map generation can be significantly enhanced (as shown in Fig.~\ref{fig:road_network}).

To address these challenge, we develop an learning-based automated system for generating lane-level simulation road networks based on street view data (as shown in Fig.~\ref{fig: framework}). By combining foundational road network data, street view image provided by open map service providers (such as Baidu Street View, \etc), and advanced learning-based lane line detection algorithms, the system accurately perceives the lane-level structure of urban road networks. It automatically extracts critical information, including road network topology, node connectivity, and turn connectivity, thereby improving the accuracy and automation level of simulation model generation. This approach not only substantially reduces the time and labor costs associated with constructing simulation road networks but also provides a scientific foundation and data support for urban traffic governance, policy formulation, and the optimization of intelligent transportation systems, laying a solid foundation for efficient and sustainable urban traffic management.

\section{Methodology}
Specifically, 
Fig.~\ref{fig: framework} illustrates the comprehensive workflow for generating simulated road networks using deep learning-based methods. The process comprises five key stages: (1) data acquisition, (2) data storage and mapping, (3) lane marking detection, (4) coordinate alignment and updating of road network data, culminating in (5) the output of vectorized road networks. Each of these stages is described in detail in the subsequent subsections.

\subsection{Collection of Road Street Views}

\begin{figure}[!t] 
    \center
    \includegraphics[width=3.4in]{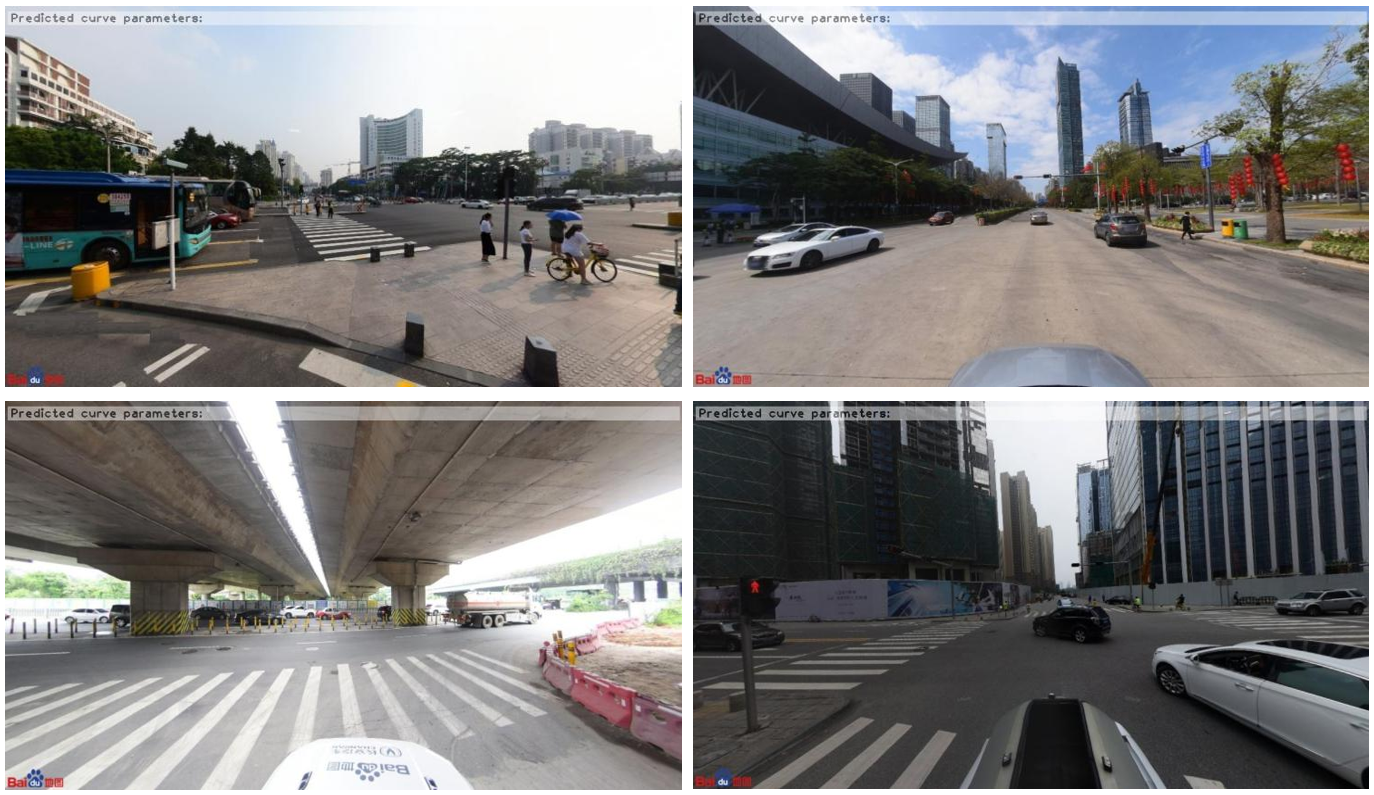}
    \caption{
    The lane detection dataset presented in this work extensively covers various corner cases, which is sufficient for effectively training high-performance lane prediction models.
    } 
    \label{fig:Lane_dataset}
\end{figure}

As shwon in Fig~\ref{fig:road_network1}, we collect real-world road street view data through open-source street view map platforms (\eg, Baidu Street View~\cite{yu2022spatio}), while meticulously recording additional attributes such as capture time and location to ensure data comprehensiveness and traceability. 
Fig.~\ref{fig:Street_View} provides a visual overview, illustrating the complete workflow for data acquisition and storage.
Based on this, a specialized dataset for lane line annotation is constructed (as shown in Fig.~\ref{fig:Lane_dataset}). To ensure the diversity and representativeness of the annotated lane line data, the dataset encompasses not only standard multi-lane street view data but also includes data from varied environments (\eg, urban arterial roads, suburban roads), different lane counts (from single to multiple lanes), diverse intersection types (\eg, cross intersections, T-junctions), various grid lines (\eg, zebra crossings, grid zones), and occlusion scenarios (\eg, vehicle occlusion, poor lighting). All data categories are rigorously curated to form a standardized multi-lane line dataset, providing high-quality training and testing data for lane line detection tasks. Furthermore, to address the prevalent issues of false positives and missed detections in current lane line detection, we implement targeted optimizations for lane line annotation, with the following specific measures:

\begin{figure}[!t]
    \center
    \includegraphics[scale=0.48]{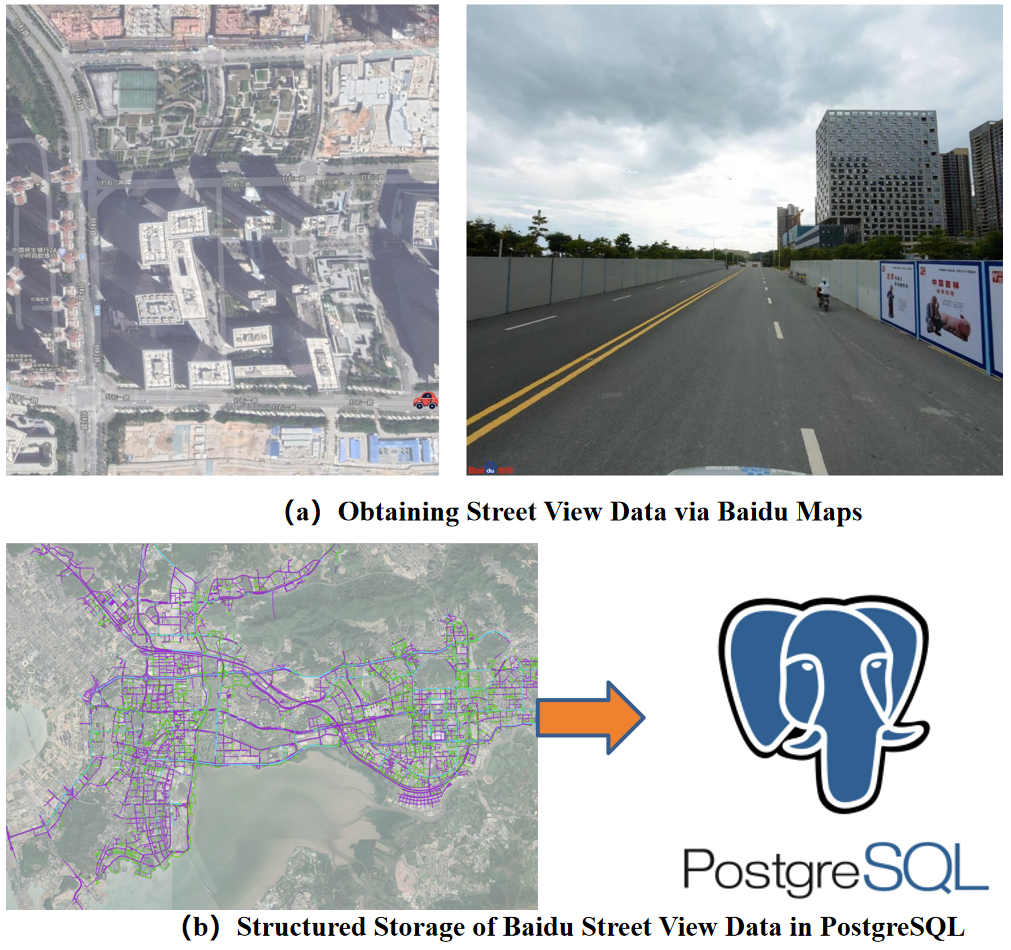}
    \caption{By utilizing a Breadth-First Search~\cite{elshewey2024orthopedic} algorithm to traverse Baidu Street View data within a specific region, the data is organized and stored in a PostgreSQL~\cite{makris2021mongodb} database.} 
    \label{fig:Street_View}
\end{figure}

\textbf{Negative Sample Set for Non-Lane Areas:} For scenarios where lane lines are absent, such as at traffic intersections or sharp turns, a dedicated negative sample set is created. By incorporating these negative samples, the model is trained to better handle situations without lane lines, enhancing its robustness in complex road conditions and reducing false detection rates.

\textbf{Standardized Annotation for Complex Grid Lines:} For various grid lines on the road surface, such as yellow grid lines in bus-only zones or no-parking areas, which are often complex and prone to confusion, we establish a detailed set of annotation standards based on the proportion of the road surface occupied by the grid and the specific location of the grid zone. These standards ensure accurate differentiation by the model, preventing prediction errors when encountering intricate grid lines and thereby improving the precision of lane line detection.

\textbf{Automatic Completion of Occluded or Worn Lane Lines:} For cases where the leftmost or rightmost lane lines are worn out and disappear, or are obscured by vehicles, an automatic completion algorithm is developed. By analyzing the geometric features and continuity of surrounding lane lines, the algorithm intelligently infers and reconstructs the occluded or missing lane lines, enhancing the dataset’s completeness and the model’s detection capabilities.

\begin{figure}[!t] 
    \center
    \includegraphics[width=3.4in]{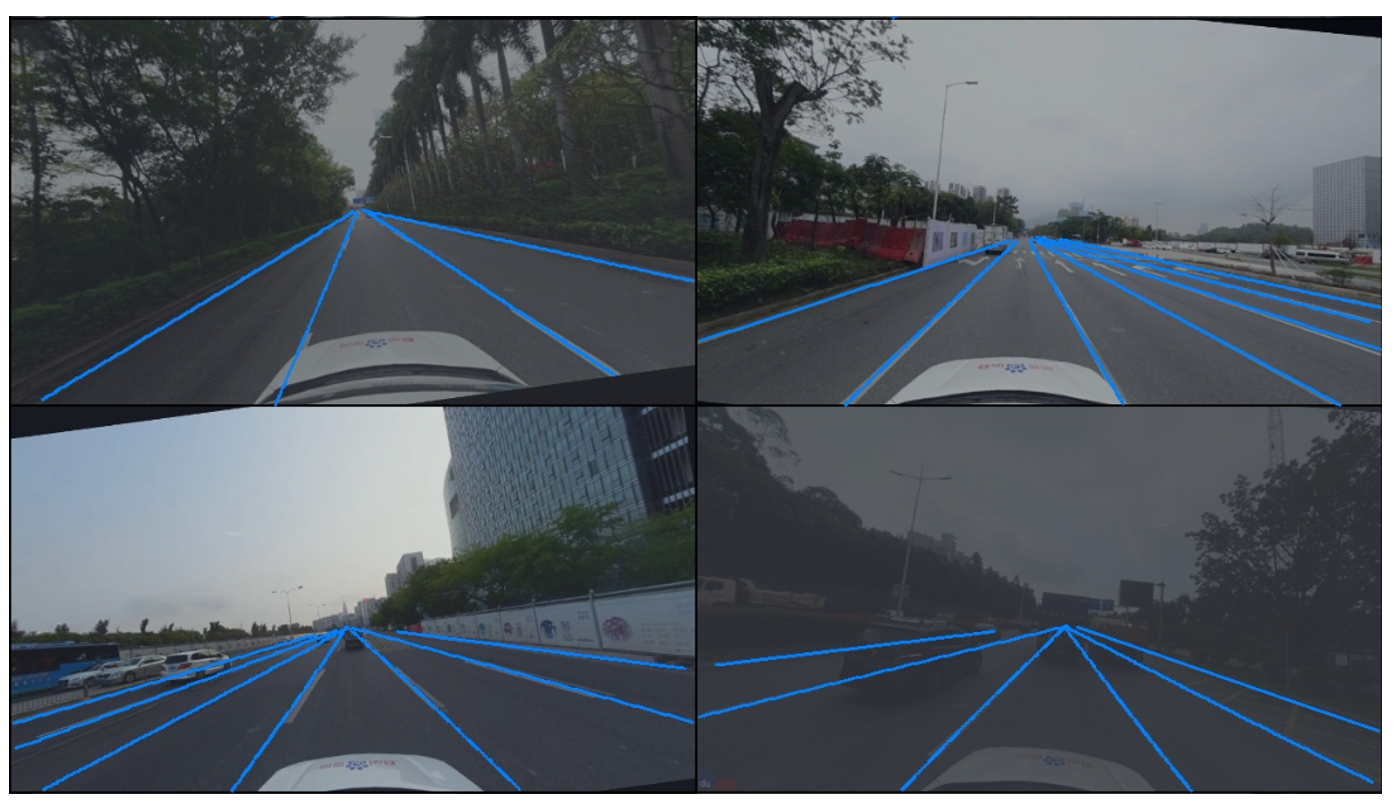}
    \caption{
    The lane detection results obtained using the Transformer-based model demonstrate superior accuracy and robustness compared to conventional CNN architectures.
    } 
    \label{fig:Lane_detection}
\end{figure} 

\textbf{Lane Line Suppression in Physically Separated Areas:} Meanwhile, in scenarios where physical separators (\eg, barriers, bollards, vehicles) exist in the middle of the road, lane line predictions on the opposite side of the obstacle may lead to anomalies. To mitigate this, lane line annotations on the side of the obstacle are suppressed during the annotation process, effectively preventing model misjudgments in physically separated areas and improving the reliability of lane line detection.

\textbf{Expanded Annotation for Multi-Lane Scenarios:} While existing public datasets typically annotate only four lanes, which is insufficient for complex urban multi-lane scenarios, the dataset we designed extends the annotation to cover more than 10 lanes based on real-world conditions. This expansion accommodates urban arterial roads, highways, and other multi-lane environments, ensuring the model can adapt to diverse urban traffic scenarios and enhancing its generalization ability in practical applications.

Through these optimization measures, we construct a high-quality, diverse lane line detection dataset, significantly improving the accuracy and robustness of lane line detection while laying a solid data foundation for subsequent simulation road network generation and urban traffic analysis~\cite{xie2025learning,xie2022online,xie2022towards,ulvi2024urban,gao2024evaluation,gao2024evaluation,raghunath2024redefining}.

\subsection{Lane Line Detection}



Conventional lane line detection~\cite{zakaria2023lane,tang2021review,zhang2021deep,you2025attention,sang2024robust,lin2024lane,yao2024building,pittner2024lanecpp,ren2025layer,lu2025lpcnet,bi2025lane,li2025road} pipelines typically involve three steps: segmenting the lane lines, aggregating the segmented results, and performing curve fitting to generate the final lane lines. However, these methods suffer from low efficiency and often overlook global contextual information during the segmentation phase, leading to suboptimal accuracy in complex scenarios. To address these issues, we employ deep learning techniques to detect the number of lanes in street view images by developing a CNN-Transformer-based network specifically designed for lane shape model prediction. This network incorporates CNN and Transformer~\cite{li2023transformer,yao2024cnn,wang2024swinurnet,cao2024novel,xu2024bridging,yang2025enhanced,zhang2025fet,cui2025cswin} components to effectively capture the local and global interactions between lane lines and their global context, enabling it to accurately model the elongated structure of lanes as well as the overall topological information of the road, such as its structure and lane count. The network directly regresses a set of parameters as its output, which not only represents the geometry shape of the lanes but also approximates road curvature and camera pose through explicit mathematical formulations, providing critical information for subsequent lane line analysis.

\begin{figure}[!t]
    \center
    \includegraphics[scale=0.34]{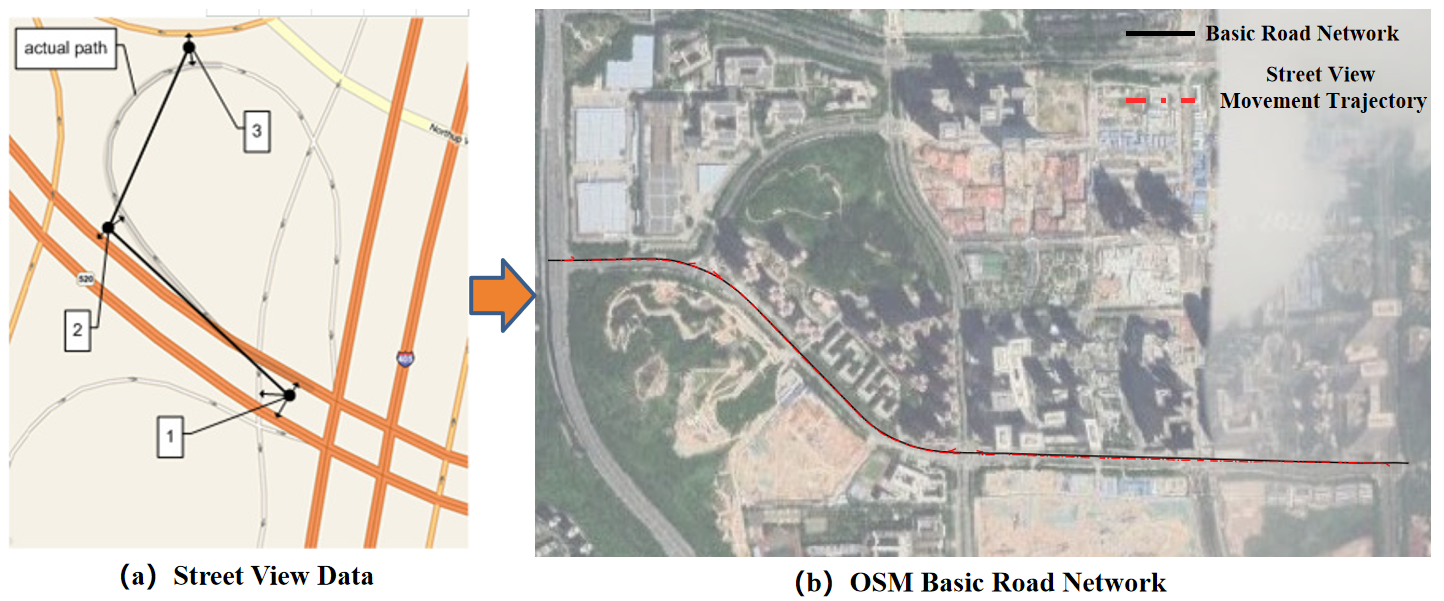}
    \caption{The trajectory similarity algorithm is implemented using the Fréchet algorithm~\cite{bringmann2021discrete}, which matches the information recognized from Baidu Street View images to the underlying vector road network.} 
    \label{fig:trajectory}
\end{figure}

Specifically, the CNN-Transformer-based network developed excels at extracting and integrating information from any pair of visual features, enabling it to effectively capture the elongated structure of lane lines and their global context. The local and global contextual information encompasses all objects within the lane line region of the image (\eg, vehicles, road signs) as well as background elements (\eg, buildings, trees along the road). The model demonstrates the capability to perform predictive analysis over large global regions, swiftly localize the range of local lane lines, and precisely delineate lane lines within smaller areas. The entire architecture is trained end-to-end, directly predicting the output while being optimized with a Hungarian loss function. The Hungarian loss~\cite{liu2021end} ensures a one-to-one unordered assignment by computing a bipartite matching between predictions and ground truth, eliminating the need for explicit non-maximum suppression (NMS)~\cite{gong2021review} commonly used in traditional methods. This design not only streamlines the detection pipeline but also enhances the model’s robustness and generalization ability in complex scenarios, offering an efficient and accurate solution for lane line detection tasks (as shown in Fig.~\ref{fig:Lane_detection}).


\begin{figure}[!t]
    \center
    \includegraphics[scale=0.47]{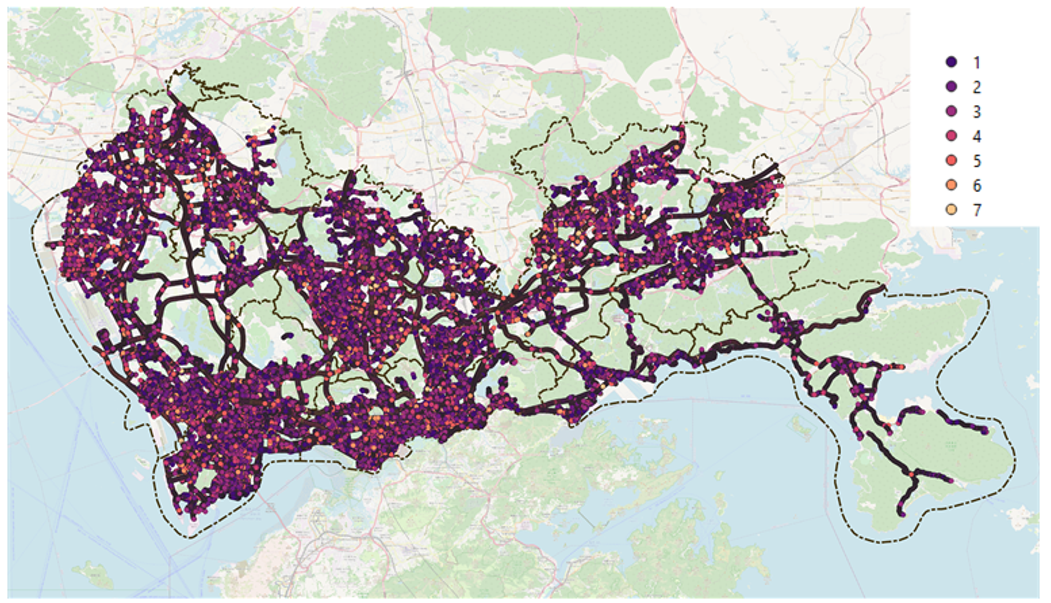}
    \caption{The road network visualization results are generated through map matching based on lane markings (covering the entire global distribution of Shenzhen's street view).} 
    \label{fig:map_matching}
\end{figure} 	

\subsection{Lane-Level Road Network Generation}

\begin{figure}[!t]
    \center
    \includegraphics[scale=0.28]{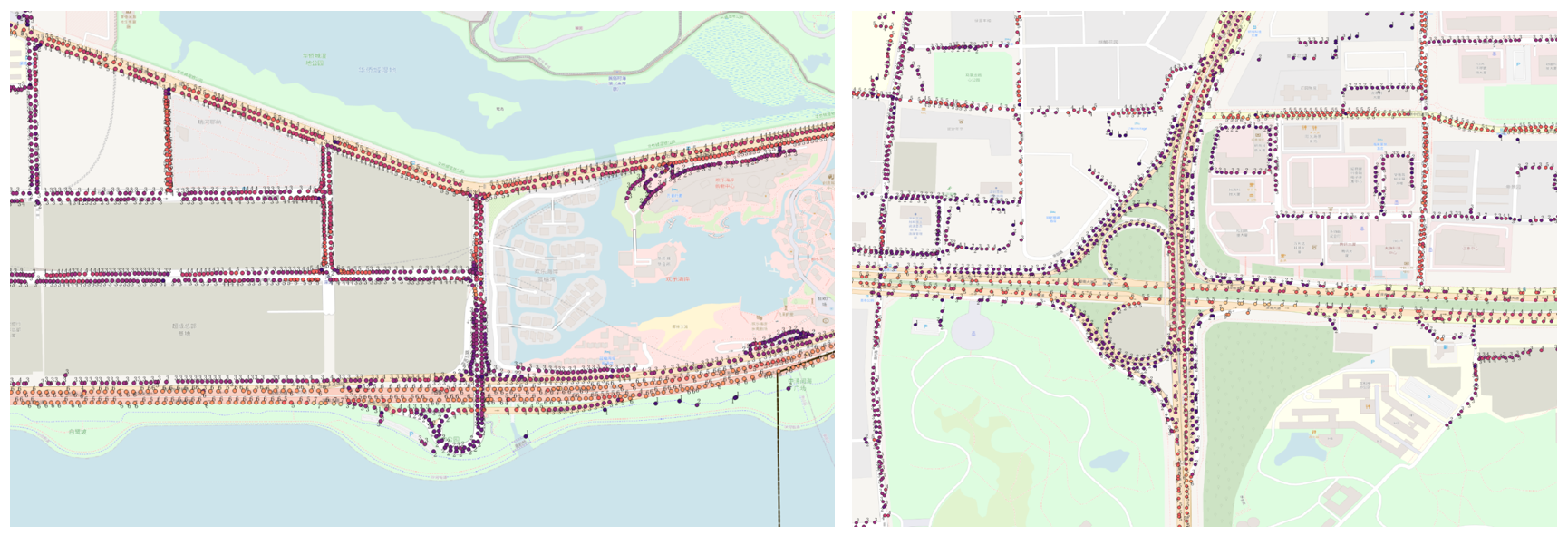}
    \caption{The distribution of the street view image layout and the prediction scenario.} 
    \label{fig:distribution}
\end{figure}
\begin{figure}[!t]
    \center
    \includegraphics[scale=0.345]{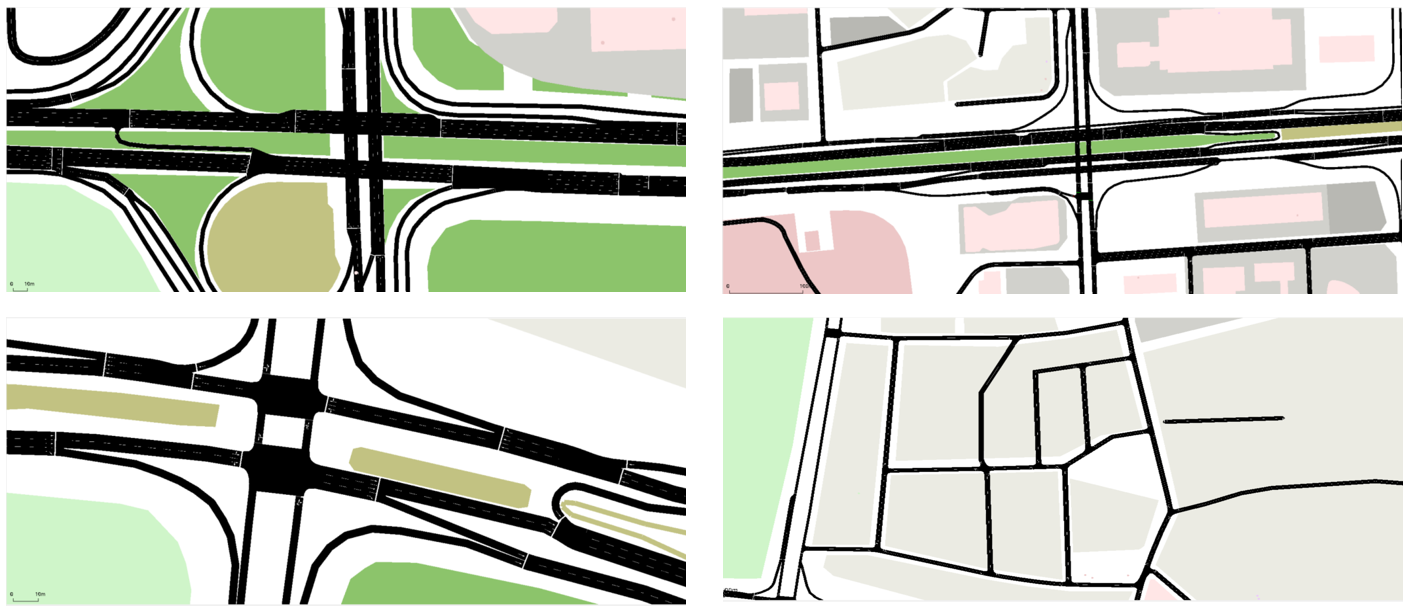}
    \caption{High-precision simulation of detailed local areas within the road network.} 
    \label{fig:simulation}
\end{figure}
\begin{figure}[!t]
    \center
    \includegraphics[scale=0.3]{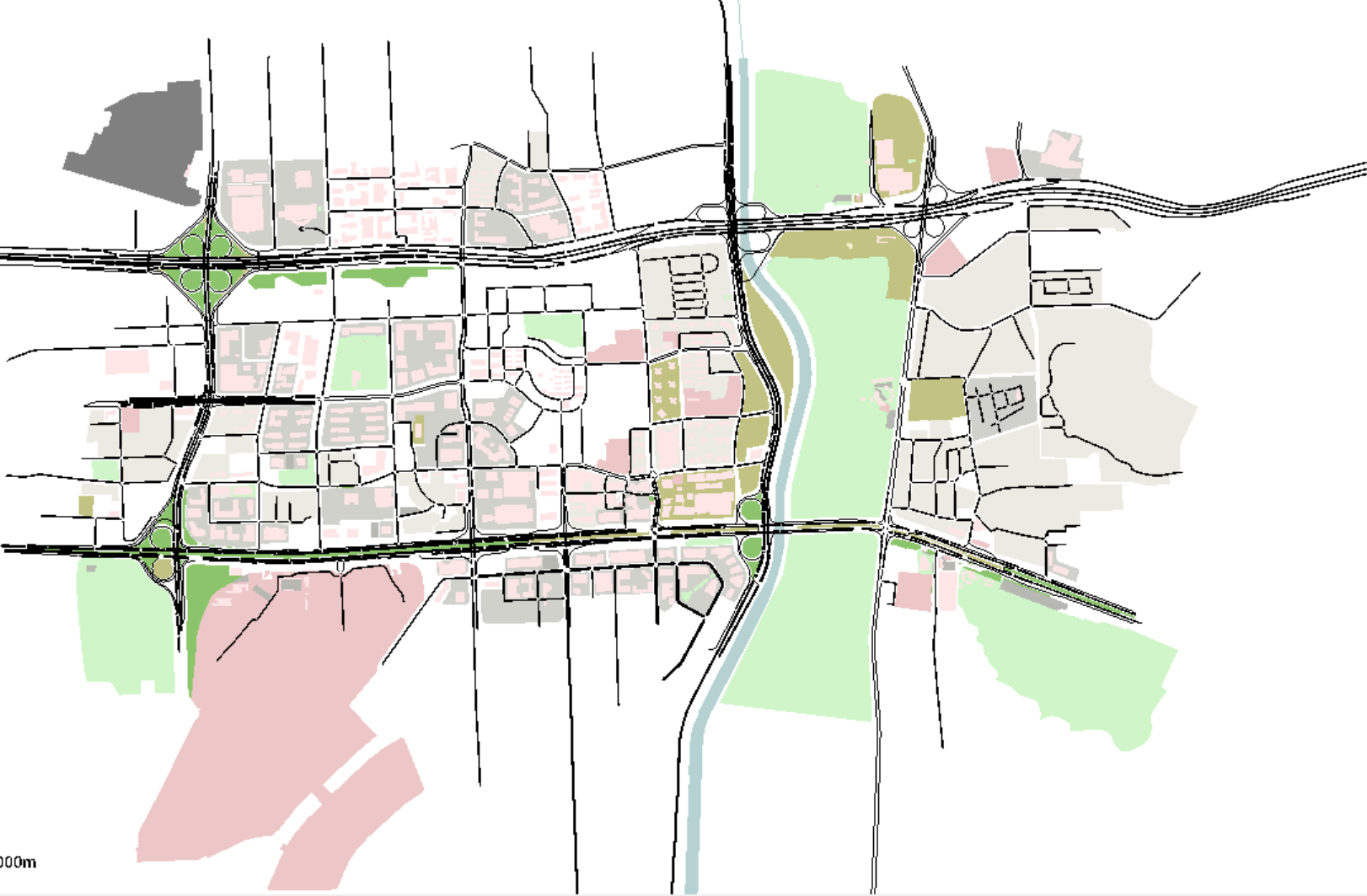}
    \caption{Complete and comprehensive  roadway network reconstruction outcomes in Shenzhen.} 
    \label{fig:simulation1}
\end{figure}

As shown in Fig.~\ref{fig:trajectory}, this study employs a map matching algorithm to seamlessly integrate the lane information predicted from street view images with the foundational road topology data collected from open-source map service platforms (\eg, Baidu Maps, OpenStreetMap), thereby developing an efficient technology for the automated generation of simulation road networks.
Fig.~\ref{fig:map_matching} presents a panoramic view of the entire road network generated for Shenzhen. Our method enables fully automated generation of urban road networks with high efficiency and speed.
Fig.~\ref{fig:distribution} provides a comparison between the original street view and the predicted road network, demonstrating that our method produces highly accurate reconstructions that closely match the real-world scenes. Additionally, Fig.~\ref{fig:simulation} showcases localized intersection results under different scenarios, further illustrating that our generated road network aligns well with the actual street conditions.
Fig.~\ref{fig:simulation1} presents the complete road network reconstruction results of Shenzhen. As can be seen from the results, the proposed method is capable of automatically generating a high-quality road network in an efficient manner. It not only significantly enhances reconstruction efficiency, but also effectively overcomes the high costs and long cycles associated with manual approaches, offering a feasible technical pathway for large-scale urban road network modeling.


This technology comprehensively generates critical components of the simulation road network, including its topological structure, node connectivity, and turn connectivity, ensuring that the resulting network is both accurate and aligned with real-world traffic scenarios. By automating this generation process, the efficiency of simulation model construction is significantly enhanced, substantially reducing the time and labor costs associated with traditional road network setup while minimizing errors introduced by manual operations. This approach not only elevates the automation level of simulation road network generation but also provides high-quality road network data for urban traffic simulation, autonomous driving testing, and the development of intelligent transportation systems, laying a solid foundation for subsequent traffic optimization and policy formulation.

\section{Conclusion}

The core strengths of our approach lie in two key aspects. First, we employ an enhanced Transformer model to detect lane line information in complex traffic scenarios, leveraging joint modeling of local and global contextual information to achieve superior performance compared to traditional methods that separately use convolution or attention mechanisms. This approach effectively captures the elongated structure of lane lines and their interactions with the surrounding environment, providing precise cross-sectional data for subsequent road network generation while improving robustness in challenging scenarios such as occlusions or intersections. Second, unlike conventional manual mapping or costly point cloud data collection methods, our approach harnesses the high accuracy and efficiency of deep learning to rapidly construct a lane-level traffic simulation network model at a lower cost. 
This model incorporates topological traffic information, road surface features, and street objects, achieving a seamless integration of street view data and road topology from open-source maps through an automated pipeline. By significantly reducing labor and time costs, this method offers efficient support for urban traffic simulation, autonomous driving testing, and intelligent transportation system development, driving advancements in the field of traffic simulation through increased automation and intelligence.


\bibliographystyle{ACM-Reference-Format}

\bibliography{sample-base}


\begin{thebibliography}{40}


\ifx \showCODEN    \undefined \def \showCODEN     #1{\unskip}     \fi
\ifx \showISBNx    \undefined \def \showISBNx     #1{\unskip}     \fi
\ifx \showISBNxiii \undefined \def \showISBNxiii  #1{\unskip}     \fi
\ifx \showISSN     \undefined \def \showISSN      #1{\unskip}     \fi
\ifx \showLCCN     \undefined \def \showLCCN      #1{\unskip}     \fi
\ifx \shownote     \undefined \def \shownote      #1{#1}          \fi
\ifx \showarticletitle \undefined \def \showarticletitle #1{#1}   \fi
\ifx \showURL      \undefined \def \showURL       {\relax}        \fi
\providecommand\bibfield[2]{#2}
\providecommand\bibinfo[2]{#2}
\providecommand\natexlab[1]{#1}
\providecommand\showeprint[2][]{arXiv:#2}

\bibitem[Bi et~al\mbox{.}(2025)]%
        {bi2025lane}
\bibfield{author}{\bibinfo{person}{Jiping Bi}, \bibinfo{person}{Yongchao Song}, \bibinfo{person}{Yahong Jiang}, \bibinfo{person}{Lijun Sun}, \bibinfo{person}{Xuan Wang}, \bibinfo{person}{Zhaowei Liu}, \bibinfo{person}{Jindong Xu}, \bibinfo{person}{Siwen Quan}, \bibinfo{person}{Zhe Dai}, {and} \bibinfo{person}{Weiqing Yan}.} \bibinfo{year}{2025}\natexlab{}.
\newblock \showarticletitle{Lane detection for autonomous driving: Comprehensive reviews, current challenges, and future predictions}.
\newblock \bibinfo{journal}{\emph{IEEE Transactions on Intelligent Transportation Systems}} (\bibinfo{year}{2025}).
\newblock


\bibitem[Bringmann et~al\mbox{.}(2021)]%
        {bringmann2021discrete}
\bibfield{author}{\bibinfo{person}{Karl Bringmann}, \bibinfo{person}{Marvin K{\"u}nnemann}, {and} \bibinfo{person}{Andr{\'e} Nusser}.} \bibinfo{year}{2021}\natexlab{}.
\newblock \showarticletitle{Discrete Fr{\'e}chet distance under translation: Conditional hardness and an improved algorithm}.
\newblock \bibinfo{journal}{\emph{ACM Transactions on Algorithms}} \bibinfo{volume}{17}, \bibinfo{number}{3} (\bibinfo{year}{2021}), \bibinfo{pages}{1--42}.
\newblock


\bibitem[Cao et~al\mbox{.}(2024)]%
        {cao2024novel}
\bibfield{author}{\bibinfo{person}{Liying Cao}, \bibinfo{person}{Miao Sun}, \bibinfo{person}{Zhicheng Yang}, \bibinfo{person}{Donghui Jiang}, \bibinfo{person}{Dongjie Yin}, {and} \bibinfo{person}{Yunpeng Duan}.} \bibinfo{year}{2024}\natexlab{}.
\newblock \showarticletitle{A novel transformer-CNN approach for predicting soil properties from LUCAS Vis-NIR spectral data}.
\newblock \bibinfo{journal}{\emph{Agronomy}} \bibinfo{volume}{14}, \bibinfo{number}{9} (\bibinfo{year}{2024}), \bibinfo{pages}{1998}.
\newblock


\bibitem[Cui et~al\mbox{.}(2025)]%
        {cui2025cswin}
\bibfield{author}{\bibinfo{person}{Jinrong Cui}, \bibinfo{person}{Youliu Zhang}, \bibinfo{person}{Hao Chen}, \bibinfo{person}{Yaoxuan Zhang}, \bibinfo{person}{Hao Cai}, \bibinfo{person}{Yu Jiang}, \bibinfo{person}{Ruijun Ma}, {and} \bibinfo{person}{Long Qi}.} \bibinfo{year}{2025}\natexlab{}.
\newblock \showarticletitle{CSWin-MBConv: A dual-network fusing CNN and Transformer for weed recognition}.
\newblock \bibinfo{journal}{\emph{European Journal of Agronomy}}  \bibinfo{volume}{164} (\bibinfo{year}{2025}), \bibinfo{pages}{127528}.
\newblock


\bibitem[Elshewey(2024)]%
        {elshewey2024orthopedic}
\bibfield{author}{\bibinfo{person}{Ahmed~M Elshewey}.} \bibinfo{year}{2024}\natexlab{}.
\newblock \showarticletitle{Orthopedic disease classification based on breadth-first search algorithm}.
\newblock \bibinfo{journal}{\emph{Scientific Reports}} \bibinfo{volume}{14}, \bibinfo{number}{1} (\bibinfo{year}{2024}), \bibinfo{pages}{23368}.
\newblock


\bibitem[Fang et~al\mbox{.}(2025)]%
        {fang2025scan}
\bibfield{author}{\bibinfo{person}{Yunping Fang}, \bibinfo{person}{Stergios-Aristoteles Mitoulis}, \bibinfo{person}{Daniel Boddice}, \bibinfo{person}{Jialiang Yu}, {and} \bibinfo{person}{Jelena Ninic}.} \bibinfo{year}{2025}\natexlab{}.
\newblock \showarticletitle{Scan-to-BIM-to-Sim: Automated reconstruction of digital and simulation models from point clouds with applications on bridges}.
\newblock \bibinfo{journal}{\emph{Results in Engineering}}  \bibinfo{volume}{25} (\bibinfo{year}{2025}), \bibinfo{pages}{104289}.
\newblock


\bibitem[Gao et~al\mbox{.}(2024)]%
        {gao2024evaluation}
\bibfield{author}{\bibinfo{person}{Sa Gao}, \bibinfo{person}{Qingsong Ran}, \bibinfo{person}{Zicheng Su}, \bibinfo{person}{Ling Wang}, \bibinfo{person}{Wanjing Ma}, {and} \bibinfo{person}{Ruochen Hao}.} \bibinfo{year}{2024}\natexlab{}.
\newblock \showarticletitle{Evaluation system for urban traffic intelligence based on travel experiences: A sentiment analysis approach}.
\newblock \bibinfo{journal}{\emph{Transportation Research Part A: Policy and Practice}}  \bibinfo{volume}{187} (\bibinfo{year}{2024}), \bibinfo{pages}{104170}.
\newblock


\bibitem[Gong et~al\mbox{.}(2021)]%
        {gong2021review}
\bibfield{author}{\bibinfo{person}{Meiling Gong}, \bibinfo{person}{Dong Wang}, \bibinfo{person}{Xiaoxia Zhao}, \bibinfo{person}{Huimin Guo}, \bibinfo{person}{Donghao Luo}, {and} \bibinfo{person}{Min Song}.} \bibinfo{year}{2021}\natexlab{}.
\newblock \showarticletitle{A review of non-maximum suppression algorithms for deep learning target detection}. In \bibinfo{booktitle}{\emph{Seventh Symposium on Novel Photoelectronic Detection Technology and Applications}}, Vol.~\bibinfo{volume}{11763}. SPIE, \bibinfo{pages}{821--828}.
\newblock


\bibitem[Li et~al\mbox{.}(2025)]%
        {li2025road}
\bibfield{author}{\bibinfo{person}{Dan Li}, \bibinfo{person}{Zan Yang}, \bibinfo{person}{Wei Nai}, \bibinfo{person}{Yidan Xing}, {and} \bibinfo{person}{Ziyu Chen}.} \bibinfo{year}{2025}\natexlab{}.
\newblock \showarticletitle{A road lane detection approach based on reformer model}.
\newblock \bibinfo{journal}{\emph{Egyptian Informatics Journal}}  \bibinfo{volume}{29} (\bibinfo{year}{2025}), \bibinfo{pages}{100625}.
\newblock


\bibitem[Li et~al\mbox{.}(2023)]%
        {li2023transformer}
\bibfield{author}{\bibinfo{person}{Yong Li}, \bibinfo{person}{Naipeng Miao}, \bibinfo{person}{Liangdi Ma}, \bibinfo{person}{Feng Shuang}, {and} \bibinfo{person}{Xingwen Huang}.} \bibinfo{year}{2023}\natexlab{}.
\newblock \showarticletitle{Transformer for object detection: Review and benchmark}.
\newblock \bibinfo{journal}{\emph{Engineering Applications of Artificial Intelligence}}  \bibinfo{volume}{126} (\bibinfo{year}{2023}), \bibinfo{pages}{107021}.
\newblock


\bibitem[Li et~al\mbox{.}(2024)]%
        {li2024choose}
\bibfield{author}{\bibinfo{person}{Yueyuan Li}, \bibinfo{person}{Wei Yuan}, \bibinfo{person}{Songan Zhang}, \bibinfo{person}{Weihao Yan}, \bibinfo{person}{Qiyuan Shen}, \bibinfo{person}{Chunxiang Wang}, {and} \bibinfo{person}{Ming Yang}.} \bibinfo{year}{2024}\natexlab{}.
\newblock \showarticletitle{Choose your simulator wisely: A review on open-source simulators for autonomous driving}.
\newblock \bibinfo{journal}{\emph{IEEE Transactions on Intelligent Vehicles}} (\bibinfo{year}{2024}).
\newblock


\bibitem[Lin et~al\mbox{.}(2024)]%
        {lin2024lane}
\bibfield{author}{\bibinfo{person}{Huei-Yung Lin}, \bibinfo{person}{Chun-Ke Chang}, {et~al\mbox{.}}} \bibinfo{year}{2024}\natexlab{}.
\newblock \showarticletitle{Lane detection networks based on deep neural networks and temporal information}.
\newblock \bibinfo{journal}{\emph{Alexandria Engineering Journal}}  \bibinfo{volume}{98} (\bibinfo{year}{2024}), \bibinfo{pages}{10--18}.
\newblock


\bibitem[Liu et~al\mbox{.}(2021)]%
        {liu2021end}
\bibfield{author}{\bibinfo{person}{Ruijin Liu}, \bibinfo{person}{Zejian Yuan}, \bibinfo{person}{Tie Liu}, {and} \bibinfo{person}{Zhiliang Xiong}.} \bibinfo{year}{2021}\natexlab{}.
\newblock \showarticletitle{End-to-end lane shape prediction with transformers}. In \bibinfo{booktitle}{\emph{Proceedings of the IEEE/CVF Winter Conference on Applications of Computer Vision}}. \bibinfo{pages}{3694--3702}.
\newblock


\bibitem[Lu et~al\mbox{.}(2025)]%
        {lu2025lpcnet}
\bibfield{author}{\bibinfo{person}{Fangfang Lu}, \bibinfo{person}{Guxue Sun}, \bibinfo{person}{Huiqun Yu}, \bibinfo{person}{Yijie Huang}, \bibinfo{person}{Tong Zhou}, {and} \bibinfo{person}{Sangyu Yao}.} \bibinfo{year}{2025}\natexlab{}.
\newblock \showarticletitle{LPCNet: End-to-end lane detection with PnP compression and condition DETR}.
\newblock \bibinfo{journal}{\emph{Displays}}  \bibinfo{volume}{87} (\bibinfo{year}{2025}), \bibinfo{pages}{102902}.
\newblock


\bibitem[Makris et~al\mbox{.}(2021)]%
        {makris2021mongodb}
\bibfield{author}{\bibinfo{person}{Antonios Makris}, \bibinfo{person}{Konstantinos Tserpes}, \bibinfo{person}{Giannis Spiliopoulos}, \bibinfo{person}{Dimitrios Zissis}, {and} \bibinfo{person}{Dimosthenis Anagnostopoulos}.} \bibinfo{year}{2021}\natexlab{}.
\newblock \showarticletitle{MongoDB Vs PostgreSQL: A comparative study on performance aspects}.
\newblock \bibinfo{journal}{\emph{GeoInformatica}}  \bibinfo{volume}{25} (\bibinfo{year}{2021}), \bibinfo{pages}{243--268}.
\newblock


\bibitem[Nguyen(2021)]%
        {nguyen2021overview}
\bibfield{author}{\bibinfo{person}{Johannes Nguyen}.} \bibinfo{year}{2021}\natexlab{}.
\newblock \showarticletitle{An overview of agent-based traffic simulators}.
\newblock \bibinfo{journal}{\emph{Transportation research interdisciplinary perspectives}}  \bibinfo{volume}{12} (\bibinfo{year}{2021}), \bibinfo{pages}{100486}.
\newblock


\bibitem[Pittner et~al\mbox{.}(2024)]%
        {pittner2024lanecpp}
\bibfield{author}{\bibinfo{person}{Maximilian Pittner}, \bibinfo{person}{Joel Janai}, {and} \bibinfo{person}{Alexandru~P Condurache}.} \bibinfo{year}{2024}\natexlab{}.
\newblock \showarticletitle{Lanecpp: Continuous 3d lane detection using physical priors}. In \bibinfo{booktitle}{\emph{Proceedings of the IEEE/CVF Conference on computer vision and pattern recognition}}. \bibinfo{pages}{10639--10648}.
\newblock


\bibitem[Quan et~al\mbox{.}(2025)]%
        {quan2025large}
\bibfield{author}{\bibinfo{person}{Hongye Quan}, \bibinfo{person}{Wanli Ni}, \bibinfo{person}{Tong Zhang}, \bibinfo{person}{Xiangyu Ye}, \bibinfo{person}{Ziyi Xie}, \bibinfo{person}{Shuai Wang}, \bibinfo{person}{Yuanwei Liu}, {and} \bibinfo{person}{Hui Song}.} \bibinfo{year}{2025}\natexlab{}.
\newblock \showarticletitle{Large language model agents for radio map generation and wireless network planning}.
\newblock \bibinfo{journal}{\emph{IEEE Networking Letters}} (\bibinfo{year}{2025}).
\newblock


\bibitem[Raghunath et~al\mbox{.}(2024)]%
        {raghunath2024redefining}
\bibfield{author}{\bibinfo{person}{KM~Karthick Raghunath}, \bibinfo{person}{C~Rohith Bhat}, \bibinfo{person}{Venkatesan~Vinoth Kumar}, \bibinfo{person}{Velmurugan~Athiyoor Kannan}, \bibinfo{person}{TR Mahesh}, \bibinfo{person}{K Manikandan}, {and} \bibinfo{person}{N Krishnamoorthy}.} \bibinfo{year}{2024}\natexlab{}.
\newblock \showarticletitle{Redefining urban traffic dynamics with TCN-FL driven traffic prediction and control strategies}.
\newblock \bibinfo{journal}{\emph{IEEE Access}}  \bibinfo{volume}{12} (\bibinfo{year}{2024}), \bibinfo{pages}{115386--115399}.
\newblock


\bibitem[Ren et~al\mbox{.}(2025)]%
        {ren2025layer}
\bibfield{author}{\bibinfo{person}{Hao Ren}, \bibinfo{person}{Mingwei Wang}, \bibinfo{person}{Yanyang Deng}, \bibinfo{person}{Wenping Li}, {and} \bibinfo{person}{Chen Liu}.} \bibinfo{year}{2025}\natexlab{}.
\newblock \showarticletitle{Layer-wise feature refinement for accurate three-dimensional lane detection with enhanced bird’s eye view transformation}.
\newblock \bibinfo{journal}{\emph{Engineering Applications of Artificial Intelligence}}  \bibinfo{volume}{152} (\bibinfo{year}{2025}), \bibinfo{pages}{110585}.
\newblock


\bibitem[Sang and Norris(2024)]%
        {sang2024robust}
\bibfield{author}{\bibinfo{person}{I-Chen Sang} {and} \bibinfo{person}{William~R Norris}.} \bibinfo{year}{2024}\natexlab{}.
\newblock \showarticletitle{A robust lane detection algorithm adaptable to challenging weather conditions}.
\newblock \bibinfo{journal}{\emph{IEEE Access}}  \bibinfo{volume}{12} (\bibinfo{year}{2024}), \bibinfo{pages}{11185--11195}.
\newblock


\bibitem[Tang et~al\mbox{.}(2021)]%
        {tang2021review}
\bibfield{author}{\bibinfo{person}{Jigang Tang}, \bibinfo{person}{Songbin Li}, {and} \bibinfo{person}{Peng Liu}.} \bibinfo{year}{2021}\natexlab{}.
\newblock \showarticletitle{A review of lane detection methods based on deep learning}.
\newblock \bibinfo{journal}{\emph{Pattern Recognition}}  \bibinfo{volume}{111} (\bibinfo{year}{2021}), \bibinfo{pages}{107623}.
\newblock


\bibitem[Tang(2022)]%
        {tang2022automatic}
\bibfield{author}{\bibinfo{person}{Yun Tang}.} \bibinfo{year}{2022}\natexlab{}.
\newblock \showarticletitle{Automatic map generation for autonomous driving system testing}.
\newblock \bibinfo{journal}{\emph{arXiv preprint arXiv:2206.09357}} (\bibinfo{year}{2022}).
\newblock


\bibitem[Teramoto et~al\mbox{.}(2025)]%
        {teramoto2025automated}
\bibfield{author}{\bibinfo{person}{Atsushi Teramoto}, \bibinfo{person}{Ayano Michiba}, \bibinfo{person}{Yuka Kiriyama}, \bibinfo{person}{Tetsuya Tsukamoto}, \bibinfo{person}{Kazuyoshi Imaizumi}, {and} \bibinfo{person}{Hiroshi Fujita}.} \bibinfo{year}{2025}\natexlab{}.
\newblock \showarticletitle{Automated description generation of cytologic findings for lung cytological images using a pretrained vision model and dual text decoders: preliminary study}.
\newblock \bibinfo{journal}{\emph{Cytopathology}} \bibinfo{volume}{36}, \bibinfo{number}{3} (\bibinfo{year}{2025}), \bibinfo{pages}{240--249}.
\newblock


\bibitem[Ulvi et~al\mbox{.}(2024)]%
        {ulvi2024urban}
\bibfield{author}{\bibinfo{person}{Hayri Ulvi}, \bibinfo{person}{Mehmet~Akif Yerlikaya}, {and} \bibinfo{person}{K{\"u}r{\c{s}}at Yildiz}.} \bibinfo{year}{2024}\natexlab{}.
\newblock \showarticletitle{Urban traffic mobility optimization model: A novel mathematical approach for predictive urban traffic analysis}.
\newblock \bibinfo{journal}{\emph{Applied Sciences}} \bibinfo{volume}{14}, \bibinfo{number}{13} (\bibinfo{year}{2024}), \bibinfo{pages}{5873}.
\newblock


\bibitem[Wang et~al\mbox{.}(2024)]%
        {wang2024swinurnet}
\bibfield{author}{\bibinfo{person}{Zhangyu Wang}, \bibinfo{person}{Zhihao Liao}, \bibinfo{person}{Bin Zhou}, \bibinfo{person}{Guizhen Yu}, {and} \bibinfo{person}{Wenwen Luo}.} \bibinfo{year}{2024}\natexlab{}.
\newblock \showarticletitle{SwinURNet: Hybrid transformer-cnn architecture for real-time unstructured road segmentation}.
\newblock \bibinfo{journal}{\emph{IEEE Transactions on Instrumentation and Measurement}} (\bibinfo{year}{2024}).
\newblock


\bibitem[Xie and Fan(2025)]%
        {xie2025learning}
\bibfield{author}{\bibinfo{person}{Liang Xie} {and} \bibinfo{person}{Songlin Fan}.} \bibinfo{year}{2025}\natexlab{}.
\newblock \showarticletitle{A Learning-based Multi-Frame Visual Feature Framework for Real-Time Driver Fatigue Detection}. In \bibinfo{booktitle}{\emph{Proceedings of the 2025 Conference of the Nations of the Americas Chapter of the Association for Computational Linguistics: Human Language Technologies (System Demonstrations)}}. \bibinfo{pages}{61--69}.
\newblock


\bibitem[Xie et~al\mbox{.}(2022a)]%
        {xie2022online}
\bibfield{author}{\bibinfo{person}{Liang Xie}, \bibinfo{person}{MengHao Hu}, {and} \bibinfo{person}{XinBei Bai}.} \bibinfo{year}{2022}\natexlab{a}.
\newblock \showarticletitle{{Online Improved Vehicle Tracking Accuracy via Unsupervised Route Generation}}. In \bibinfo{booktitle}{\emph{IEEE 34th International Conference on Tools with Artificial Intelligence}}. IEEE, \bibinfo{pages}{788--792}.
\newblock


\bibitem[Xie et~al\mbox{.}(2022b)]%
        {xie2022towards}
\bibfield{author}{\bibinfo{person}{Liang Xie}, \bibinfo{person}{MengHao Hu}, {and} \bibinfo{person}{XinBei Bai}.} \bibinfo{year}{2022}\natexlab{b}.
\newblock \showarticletitle{{Towards Hardware-Friendly and Robust Facial Landmark Detection Method}}. In \bibinfo{booktitle}{\emph{International Conference on Neural Information Processing}}. Springer, \bibinfo{pages}{432--444}.
\newblock


\bibitem[Xu et~al\mbox{.}(2024)]%
        {xu2024bridging}
\bibfield{author}{\bibinfo{person}{Fulin Xu}, \bibinfo{person}{Shaohui Mei}, \bibinfo{person}{Ge Zhang}, \bibinfo{person}{Nan Wang}, {and} \bibinfo{person}{Qian Du}.} \bibinfo{year}{2024}\natexlab{}.
\newblock \showarticletitle{Bridging CNN and transformer with cross-attention fusion network for hyperspectral image classification}.
\newblock \bibinfo{journal}{\emph{IEEE Transactions on Geoscience and Remote Sensing}}  \bibinfo{volume}{62} (\bibinfo{year}{2024}), \bibinfo{pages}{1--14}.
\newblock


\bibitem[Yang et~al\mbox{.}(2025)]%
        {yang2025enhanced}
\bibfield{author}{\bibinfo{person}{Junjie Yang}, \bibinfo{person}{Haibo Wan}, {and} \bibinfo{person}{Zhihai Shang}.} \bibinfo{year}{2025}\natexlab{}.
\newblock \showarticletitle{Enhanced hybrid CNN and transformer network for remote sensing image change detection}.
\newblock \bibinfo{journal}{\emph{Scientific Reports}} \bibinfo{volume}{15}, \bibinfo{number}{1} (\bibinfo{year}{2025}), \bibinfo{pages}{10161}.
\newblock


\bibitem[Yao et~al\mbox{.}(2024b)]%
        {yao2024building}
\bibfield{author}{\bibinfo{person}{Jiawei Yao}, \bibinfo{person}{Xiaochao Pan}, \bibinfo{person}{Tong Wu}, {and} \bibinfo{person}{Xiaofeng Zhang}.} \bibinfo{year}{2024}\natexlab{b}.
\newblock \showarticletitle{Building lane-level maps from aerial images}. In \bibinfo{booktitle}{\emph{ICASSP 2024-2024 IEEE International Conference on Acoustics, Speech and Signal Processing (ICASSP)}}. IEEE, \bibinfo{pages}{3890--3894}.
\newblock


\bibitem[Yao et~al\mbox{.}(2024a)]%
        {yao2024cnn}
\bibfield{author}{\bibinfo{person}{Wenjian Yao}, \bibinfo{person}{Jiajun Bai}, \bibinfo{person}{Wei Liao}, \bibinfo{person}{Yuheng Chen}, \bibinfo{person}{Mengjuan Liu}, {and} \bibinfo{person}{Yao Xie}.} \bibinfo{year}{2024}\natexlab{a}.
\newblock \showarticletitle{From cnn to transformer: A review of medical image segmentation models}.
\newblock \bibinfo{journal}{\emph{Journal of Imaging Informatics in Medicine}} \bibinfo{volume}{37}, \bibinfo{number}{4} (\bibinfo{year}{2024}), \bibinfo{pages}{1529--1547}.
\newblock


\bibitem[You et~al\mbox{.}(2025)]%
        {you2025attention}
\bibfield{author}{\bibinfo{person}{Feng You}, \bibinfo{person}{Yi Xie}, \bibinfo{person}{Siyi Zhang}, \bibinfo{person}{Hao Chen}, \bibinfo{person}{Haiwei Wang}, \bibinfo{person}{Wei Zhang}, {and} \bibinfo{person}{Jianrong Liu}.} \bibinfo{year}{2025}\natexlab{}.
\newblock \showarticletitle{Attention based network for real-time road drivable area, lane line detection and scene identification}.
\newblock \bibinfo{journal}{\emph{Engineering Applications of Artificial Intelligence}}  \bibinfo{volume}{160} (\bibinfo{year}{2025}), \bibinfo{pages}{111781}.
\newblock


\bibitem[Yu et~al\mbox{.}(2022)]%
        {yu2022spatio}
\bibfield{author}{\bibinfo{person}{Xinyang Yu}, \bibinfo{person}{Younggu Her}, \bibinfo{person}{Wenqian Huo}, \bibinfo{person}{Guowei Chen}, {and} \bibinfo{person}{Wei Qi}.} \bibinfo{year}{2022}\natexlab{}.
\newblock \showarticletitle{Spatio-temporal monitoring of urban street-side vegetation greenery using Baidu Street View images}.
\newblock \bibinfo{journal}{\emph{Urban Forestry \& Urban Greening}}  \bibinfo{volume}{73} (\bibinfo{year}{2022}), \bibinfo{pages}{127617}.
\newblock


\bibitem[Zakaria et~al\mbox{.}(2023)]%
        {zakaria2023lane}
\bibfield{author}{\bibinfo{person}{Noor~Jannah Zakaria}, \bibinfo{person}{Mohd~Ibrahim Shapiai}, \bibinfo{person}{Rasli Abd~Ghani}, \bibinfo{person}{Mohd Najib~Mohd Yassin}, \bibinfo{person}{Mohd~Zamri Ibrahim}, {and} \bibinfo{person}{Nurbaiti Wahid}.} \bibinfo{year}{2023}\natexlab{}.
\newblock \showarticletitle{Lane detection in autonomous vehicles: A systematic review}.
\newblock \bibinfo{journal}{\emph{IEEE access}}  \bibinfo{volume}{11} (\bibinfo{year}{2023}), \bibinfo{pages}{3729--3765}.
\newblock


\bibitem[Zha et~al\mbox{.}(2023)]%
        {zha2023survey}
\bibfield{author}{\bibinfo{person}{Yunfei Zha}, \bibinfo{person}{Jianxian Deng}, \bibinfo{person}{Yinyuan Qiu}, \bibinfo{person}{Kun Zhang}, {and} \bibinfo{person}{Yanyan Wang}.} \bibinfo{year}{2023}\natexlab{}.
\newblock \showarticletitle{A survey of intelligent driving vehicle trajectory tracking based on vehicle dynamics}.
\newblock \bibinfo{journal}{\emph{SAE International journal of vehicle dynamics, stability, and NVH}} \bibinfo{volume}{7}, \bibinfo{number}{10-07-02-0014} (\bibinfo{year}{2023}), \bibinfo{pages}{221--248}.
\newblock


\bibitem[Zhang et~al\mbox{.}(2025)]%
        {zhang2025fet}
\bibfield{author}{\bibinfo{person}{Huaikun Zhang}, \bibinfo{person}{Jing Lian}, {and} \bibinfo{person}{Yide Ma}.} \bibinfo{year}{2025}\natexlab{}.
\newblock \showarticletitle{FET-UNet: Merging CNN and transformer architectures for superior breast ultrasound image segmentation}.
\newblock \bibinfo{journal}{\emph{Physica Medica}}  \bibinfo{volume}{133} (\bibinfo{year}{2025}), \bibinfo{pages}{104969}.
\newblock


\bibitem[Zhang et~al\mbox{.}(2021)]%
        {zhang2021deep}
\bibfield{author}{\bibinfo{person}{Youcheng Zhang}, \bibinfo{person}{Zongqing Lu}, \bibinfo{person}{Xuechen Zhang}, \bibinfo{person}{Jing-Hao Xue}, {and} \bibinfo{person}{Qingmin Liao}.} \bibinfo{year}{2021}\natexlab{}.
\newblock \showarticletitle{Deep learning in lane marking detection: A survey}.
\newblock \bibinfo{journal}{\emph{IEEE Transactions on Intelligent Transportation Systems}} \bibinfo{volume}{23}, \bibinfo{number}{7} (\bibinfo{year}{2021}), \bibinfo{pages}{5976--5992}.
\newblock


\bibitem[Zhong et~al\mbox{.}(2023)]%
        {zhong2023guided}
\bibfield{author}{\bibinfo{person}{Ziyuan Zhong}, \bibinfo{person}{Davis Rempe}, \bibinfo{person}{Danfei Xu}, \bibinfo{person}{Yuxiao Chen}, \bibinfo{person}{Sushant Veer}, \bibinfo{person}{Tong Che}, \bibinfo{person}{Baishakhi Ray}, {and} \bibinfo{person}{Marco Pavone}.} \bibinfo{year}{2023}\natexlab{}.
\newblock \showarticletitle{Guided conditional diffusion for controllable traffic simulation}. In \bibinfo{booktitle}{\emph{IEEE International Conference on Robotics and Automation}}. IEEE, \bibinfo{pages}{3560--3566}.
\newblock


\end{thebibliography}
\end{document}